\documentclass[twocolumn,twoside,slac_two]{revtex4}
\usepackage{graphicx}
\usepackage{fancyhdr}
\begin{document}

\hspace{5.2in}\mbox{FERMILAB-CONF-07-615-E}
\title{Rare D Decays}

%

\author{B.C.K. Casey}
\affiliation{Fermilab, Batavia, IL 60510, USA}

\begin{abstract}
We discuss several recent measurements of rare charmed hadron decays. 
Focus is placed on radiative and annihilation topologies highlighting 
their sensitivity to new physics and pointing out the strengths and 
weaknesses of different channels. We compare the different measurement
 techniques employed at fixed target and $e^+e^-$ dedicated charm 
experiments, B-factories, and the Tevatron experiments.  Comparisons
 are also made to similar topologies in the beauty, strange, and top systems 
where appropriate.
\end{abstract}

\maketitle

\thispagestyle{fancy}


\section{Introduction}
Many extensions of the standard model (SM) predict anomalous effects in rare
decays of beauty, charmed, and strange hadrons that could significantly alter
 their decay rate with respect to SM expectations.  
In $B$ meson decays, the experimental sensitivity has reached the
 SM expected rates for many rare processes. In contrast, GIM suppression~\cite{bib:gim} in $D$ meson
 decays is significantly stronger and the SM branching fractions, 
in the case of radiative $D$ meson decays, are expected
 to be as low as $10^{-9}$~\cite{bib:pakvasa,bib:fajer}.  
This leaves a large window of opportunity still available to search for new physics in charm decays.

Annihilation topologies of charged mesons can be used to probe new charged 
current phenomena that would appear at tree level such such as models with
 charged Higgs bosons~\cite{bib:akeroyd}. Here, the advantage is the SM decay 
rate can be precisely calculated and the rates are experimentally accessible. 
 Given apriori knowledge of the decay constants and CKM elements, measurements 
of these processes can place strict bounds on new phenomena.

As a third generation particle, sizable corrections are expected to $B^+$ 
annihilation in SUSY models with high $\tan\beta$~\cite{bib:carena}.  
Sensitivity to new physics in these decays are currently limited by statistics 
but will eventually be limited by errors in $V_{ub}$ and $f_B$.
As a second generation particle, the corrections are expected to be less evident
 in $D^+_{(s)}$ annihilation~\cite{bib:akeroyd}.  However, statistics
 are now available to make precision measurements of both 
$D_s\rightarrow \tau\nu$ and $D_s\rightarrow \mu\nu$. 
The ratio of these channels can then provide an extremely clean test 
for models that do not preserve lepton universality.

Radiative meson decay and annihilation of neural mesons are sensitive 
to tree level neutral current phenomena or almost any new particle that 
can interact at higher order through penguin or box diagrams.  The SM rate 
is absent at tree level and thus always suppressed.  The precision to which 
the SM rate can be calculated varies drastically depending on generation and topology.
For radiative beauty transitions such as $b\rightarrow s\gamma$ precision measurements 
and calculations are available~\cite{bib:becker}. For strange meson transitions 
such as $K_L\rightarrow \pi^0\nu{\bar\nu}$, precision calculations exist and 
the SM rates are expected to be accessible in the next generation of kaon 
experiments.

For radiative charmed hadron decays
 such as $c\rightarrow u l^+l^-$, the SM rate is extremely difficult to estimate.
  However, given the present level of experimental sensitivity, the errors in 
imperfect cancellation through the GIM mechanism can be ignored and we can 
essentially treat these decays as forbidden.  Thus at the current level of sensitivity,
 any signal in the charm sector would unambiguously signal new physics.  
This relation between current experimental sensitivity and SM expectations is also 
true for annihilation of neutral $B$ and $D$ mesons.  In this situation, 
any improvement of experimental limits allows us to place further constraints on new phenomena.

\section{Experimental Environments}
Results are available from a extremely diverse set of experiments.  The cleanest environment 
is $e^+e^-$ at charm threshold such as CLEO-c.  Here, beam constraints are a powerful tool 
in background reduction and CESR has now delivered enough luminosity at particular 
resonances to allow for competitive studies of rare decays.

Some of the largest charm samples are available at the B-factory experiments 
Belle and BaBar where the direct charm production cross section is similar 
to the $\Upsilon$(4S) production cross section and all species of charmed hadrons 
are available in the same data set.  Since the final state is dominated pions, the 
excellent particle ID capabilities of these experiments greatly reduces the 
combinatorial background in $D$ and $\Lambda_c$ decays where either multiple 
kaons or protons are present. While not at threshold, the 
isolation of direct charm production still allows for strong background reduction 
through global event variables such as the total and missing energy in the event.

Results are available from many fixed target experiments conducted in the last 
decade at Fermilab with the best limits on rare decays coming from FOCUS~\cite{bib:focus} that set 
the bar for the current experiments.  The large boost 
and excellent vertexing capabilities of these experiments led to large high purity samples 
of all charm species.  While these data sets have now been surpassed by other experiments, they 
still remind us of opportunities that will become available at LHCb or possibly 
future dedicated fixed target charm experiments at Fermilab~\cite{bib:schwartz} that 
will have similar analysis strategies but much larger data sets.

Run II of the Fermilab Tevatron has brought the study of rare charm to the energy 
frontier experiments D\O and CDF.  Here again all species are available and
the enormous production cross sections more than compensate for the lower luminosity.  
However for rare decays, the large backgrounds lead to stringent limitations 
on the channels available for study and to date, only final states containing dimuons have been considered.

\section{Annihilation}

\subsection{Charged Meson Annihilation}

New results are available this summer from the Belle Collaboration on the 
decay $D_s\rightarrow \mu\nu$~\cite{bib:belle}.  Belle reports
$${\cal B}(D_s\rightarrow \mu\nu) = (6.44\pm 0.76 \pm 0.52)\times 10^{-3}.$$
Combining this measurement with PDG'06~\cite{bib:pdg} 
and BaBar~\cite{bib:babar1} and CLEO-c~\cite{bib:cleo1} measurements from 
2007 indicate an experimental sensitivity on the order of $8\%$ in this branching 
fraction and indicate that the ratio of experimental measurement to theoretical 
prediction for $D_s\rightarrow \tau\nu/D_s\rightarrow \mu\nu$ 
can now be determined to roughly $15\%$.  This can be compared to the experiment
 to theory ratio in $B^+\rightarrow \tau\nu$ that is measured to a precision 
of about $44\%$, the recently observed Belle measurement of 
$B\rightarrow D^*\tau\nu$~\cite{bib:belle2} that has a precision of about $30\%$, or the recent 
measurement of $t\bar{t}$ production cross section with 
$t\rightarrow b\tau\nu$~\cite{bib:dzero} that also has a precision of about $30\%$. 
So while the $c\rightarrow\tau$ transition is not expected to have contributions
 as large as those in the top and $b$ systems, it makes up for it with both 
experimental and theoretical precision.

\subsection{Neutral Meson Annihilation}

\begin{figure}[h]
\centering
\includegraphics[width=70mm]{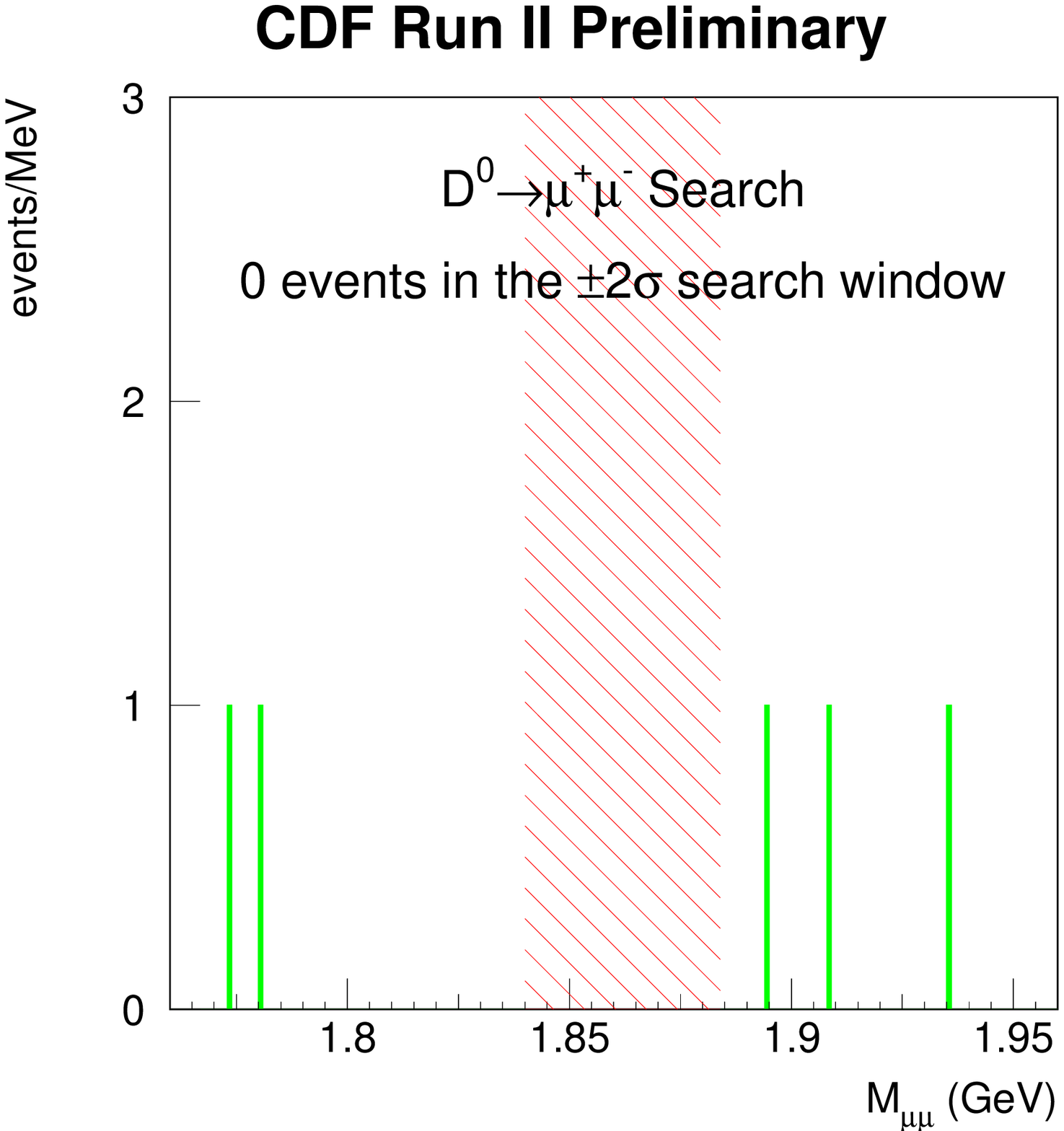}
\includegraphics[width=70mm]{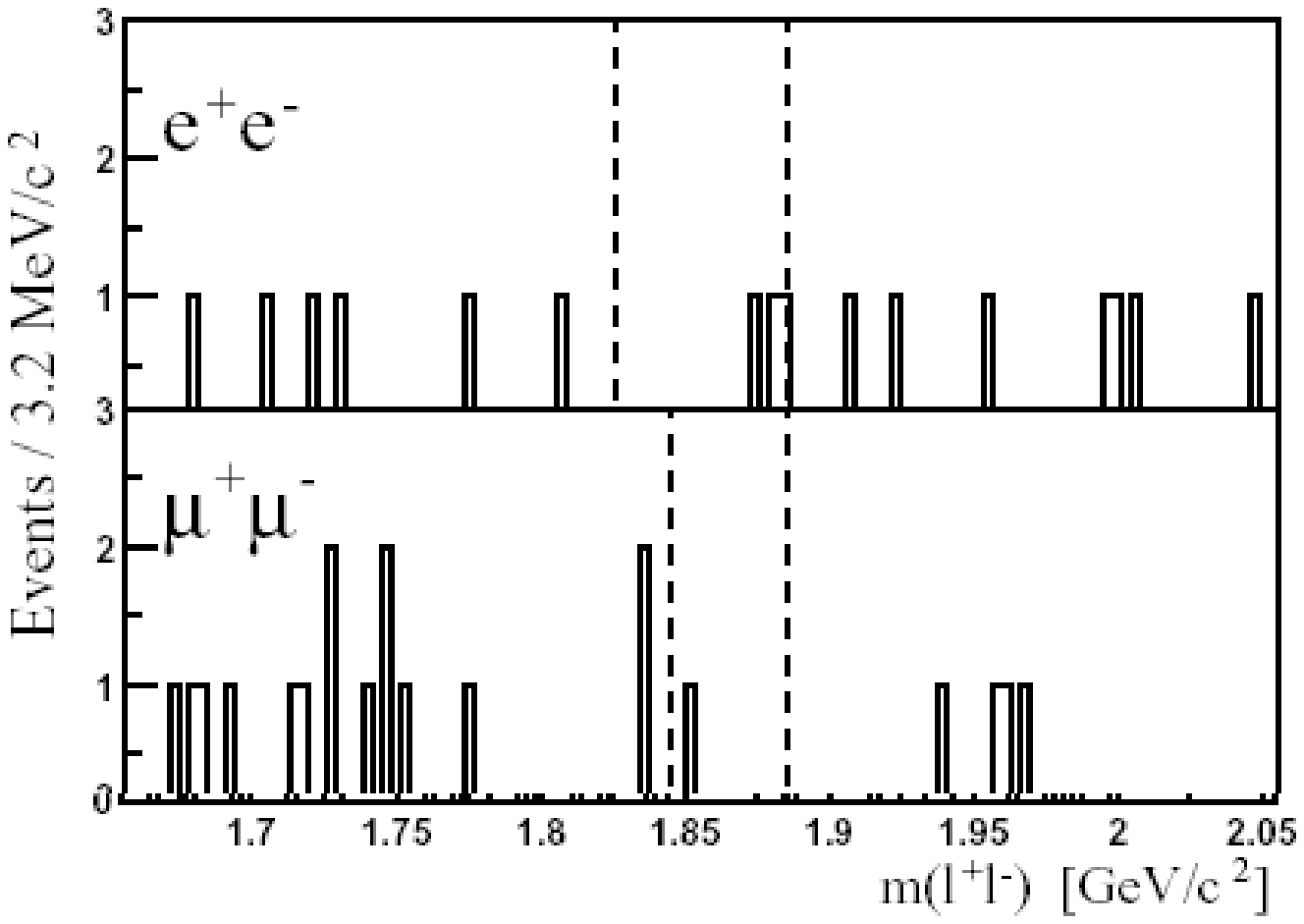}
\caption{Dilepton invariant mass distributions from CDF in the dimuon channel (above) and BaBar in the dimuon and dielectron channels in the $D^0\rightarrow l^+l^-$ analyses.} \label{fig:mll}
\end{figure}

The best limits on $D^0$ annihilation have recently been reported by the CDF~\cite{bib:cdf}
 and BaBar~\cite{bib:babar2} collaborations.  For normalization purposes, both analyses first 
reconstruct a large sample of $D^*$ tagged $D^0\rightarrow \pi^+\pi^-$ decays.
CDF reconstructs about $1.4$k $D^0\rightarrow \pi^+\pi^-$ decays in a 
65 pb$^{-1}$ data sample while BaBar reconstructs greater than $7$k
 $D^0\rightarrow \pi^+\pi^-$ decays in a 122 fb$^{-1}$ data sample.  The CDF
 analysis focuses on the dimuon final state while BaBar reconstructs 
both $\mu^+\mu^-$ and $e^+e^-$.  The possible peaking background from double 
misidentification of $D^0\rightarrow\pi^+\pi^-$ as $\mu^+\mu^-$ is studied 
using large samples of $D^*$ tagged $D^0\rightarrow K\pi$ decays.  CDF sets a $90\%$ CL upper limit of
$${\cal B}(D^0\rightarrow \mu^+\mu^-) < 2.5 \times 10^{-6},$$
while BaBar sets $90\%$ CL upper limits of
$${\cal B}(D^0\rightarrow \mu^+\mu^-) < 1.3 \times 10^{-6},$$
$${\cal B}(D^0\rightarrow e^+e^-) < 1.2 \times 10^{-6}.$$
The final dilepton invariant mass distributions are shown in Fig.~\ref{fig:mll}.

\section{Radiative Decay}

The first radiative charm decay to be observed is the decay 
$D_s\rightarrow \phi\gamma$~\cite{bib:belle3} where 
Belle measures
$${\cal B}(D_s\rightarrow\phi\gamma) = (2.6^{+0.70}_{-0.61} \hskip 1mm^{+0.15}_{-0.17})\times 10^{-5}.$$
This is a beautiful measurement where many of the peaking backgrounds
 such as $\phi\pi^0$ and $\phi\eta$ could not be constrained using
 previous information and thus were concurrently measured along with 
the $\phi\gamma$ final state.

This result is also an excellent example of the inherent problems caused by 
long distance effects in the charm system.  In the above channel, one can not 
distinguish between the quark level $c\bar{u}\rightarrow s\bar{s}\gamma$ 
transition and long distance rescattering of intermediate $D^0\rightarrow \phi \rho$ 
or $\phi\omega$ transitions into the $\phi\gamma$ final state.  Since the rate of 
these final state interactions can not be calculated with acceptable precision, 
no limits can be placed on new phenomena using the above channel~\cite{bib:pakvasa2}.

This situation can be solved by moving from two-body to three-body radiative
 decays where the extra kinematic information in the final state allows for 
a separation of long distance and short distance components~\cite{bib:pakvasa,bib:fajer}.  For instance 
in the decay $D^+\rightarrow\pi^+\mu^+\mu^-$  the long distance rescattering 
of $\phi\rightarrow\mu^+\mu^-$ can be extracted from the dimuon invariant mass spectra.
Since the short distance component is expected to be three orders of magnitude
 below the long distance component, any excess in the dimuon mass spectra away
 from the $\phi$ resonance would clearly indicate new physics.

The best limits on the $c\rightarrow u l^+l^-$ transition come from CLEO-c~\cite{bib:cleo2}, BaBar~\cite{bib:babar3},
 and D\O~\cite{bib:dzero2}.  The CLEO-c analysis is based on a data sample of 281 pb$^{-1}$ 
recorded at the $\psi(3770)$ resonance.  
The excellent calorimetry at CLEO-c leads to a focus on the $\pi^+e^+e^-$ final state.
The BaBar analysis is based on $281$ fb$^{-1}$.  The combination of powerful hadron 
and lepton ID systems allow BaBar to search for both dimuon and dielectron final 
states of $D^+$, $D_s$, and $\Lambda_c$.  The D\O analysis is based on a 1.3 fb${^-1}$ 
data sample.  The excellent dimon trigger system leads to a focus on 
the dimuon final state.  Since the background reduction techniques 
rely heavily on secondary verticies reconstructed away from the interaction point, 
focus is placed on the $D^+$ meson rather than the $D_s$ or $\Lambda_c$ 
due to their shorter lifetimes.

\begin{figure}[h]
\centering
\includegraphics[width=60mm]{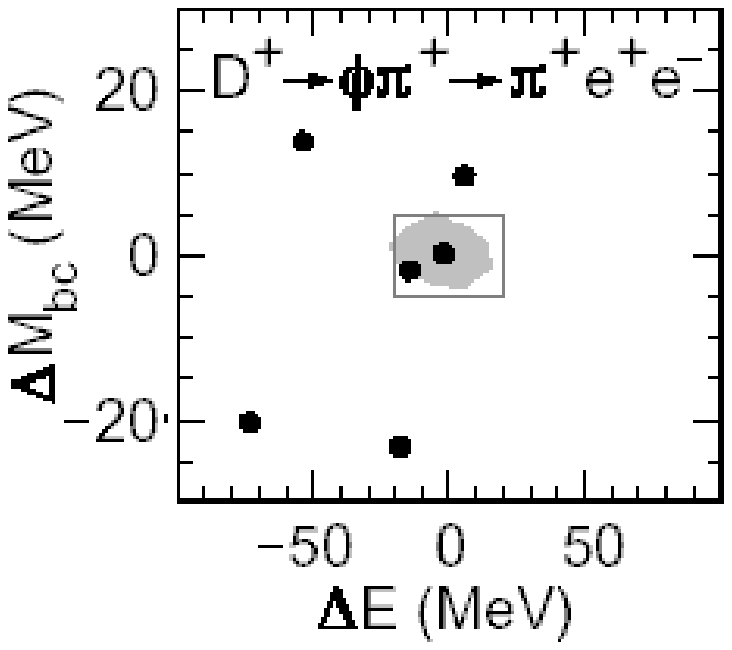}
\includegraphics[width=60mm]{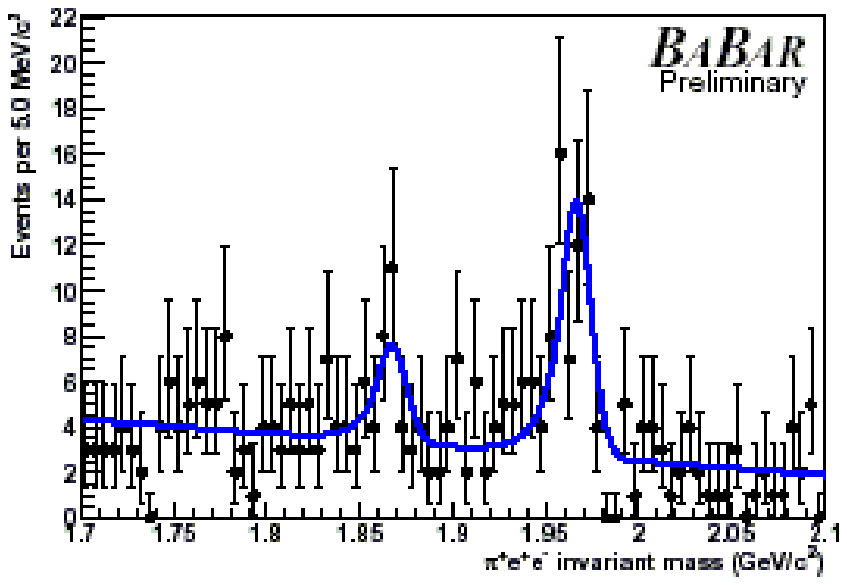}
\includegraphics[width=60mm]{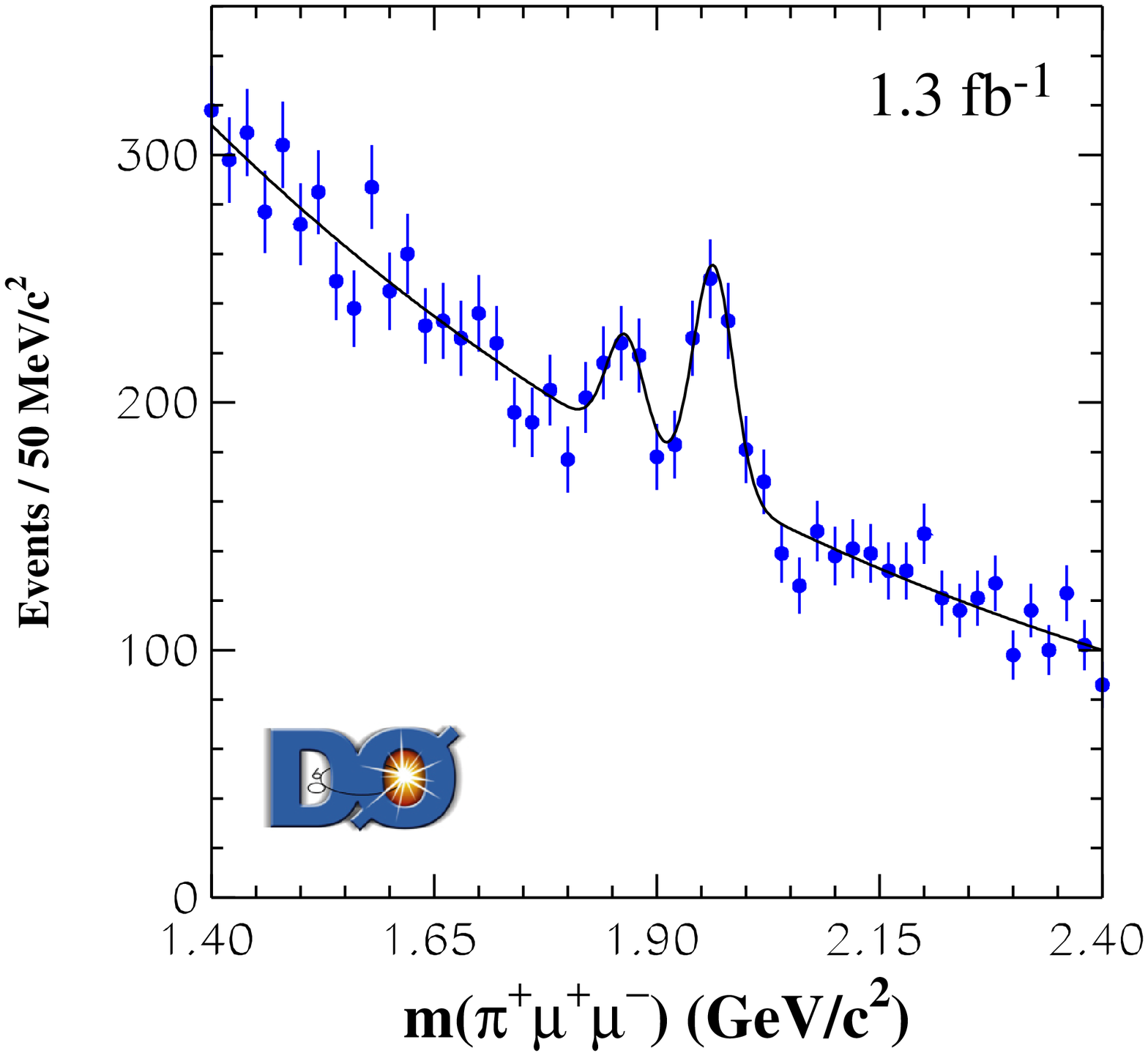}
\caption{Results of the search for $D^+\rightarrow \pi\phi \rightarrow \pi l^+l^-$.  The top figure is the beam constrained mass versus beam energy difference from CLEO in the dielectron channel.  The middle figure is the $\pi e^+e^-$ invariant mass from BaBar.  The lower figure is the $\pi\mu^+\mu^-$ invariant mass distribution from D\O .}\label{fig:phipi}
\end{figure}

As a first step, all three collaborations attempt to establish the long 
distance component $D^+\rightarrow\phi\pi^+\rightarrow l^+l^-\pi^+$ by requiring
 the dilepton invariant mass be consistent with a $\phi$. The results are shown in Fig.~\ref{fig:phipi}.
 CLEO-c finds two events 
with an expected background of $0.04$ events.  BaBar sees 19 signal events over a background of about 30 events.  D\O  sees 115 signal events over a background
 of roughly 850 events.  The differences in the environments are clearly seen in 
these yield and background comparisons.  The three collaborations measure
\begin{eqnarray*}
\lefteqn{{\cal B}(D^+\rightarrow\phi\pi^+\rightarrow e^+e^- \pi^+) =} \\
& (2.7^{+3.6}_{-1.8}\pm 0.2)\times 10^{-6} \hskip 1mm\rm (CLEO),\\
& (2.7^{+3.6}_{-1.8}\pm 0.2)\times 10^{-6} \hskip 1mm\rm (BaBar),\\
\lefteqn{{\cal B}(D^+\rightarrow\phi\pi^+\rightarrow \mu^+\mu^- \pi^+) =}\\
&(1.8 \pm 0.5\pm 0.6)\times 10^{-6} \hskip 1mm\rm (D\slash{O}).
\end{eqnarray*}


With the long distance contribution established, each analysis proceeds to 
search for the short distance $c\rightarrow u l^+l^-$ transition by looking 
for an excess of events away from the $\phi$ resonance.  CLEO-c takes advantage 
of beam constrained variables and detector hermiticity to specifically veto the
 dominant background of two semileptonic $D$ decays and arrives at a $90\%$ CL upper limit of
$${\cal B}(D^+\rightarrow \pi^+ e^+e^-) < 7.4 \times 10^{-6} \hskip 1mm\rm (CLEO).$$

The BaBar analysis requires high momentum $D$ candidates consistent with 
direct $c\bar{c}$ production to remove backgrounds from semileptonic $B$
 decay and then also relies on hermiticity to remove backgrounds from
 two semileptonic charm decays.  Using $\Lambda_c$ decays easily distinguished using particle ID, 
they set the best $90\%$ CL 
upper limit in the dielectron channel of
$${\cal B}(\Lambda_c\rightarrow p e^+e^-) < 3.6\times 10^{-6} \hskip 1mm\rm (BaBar).$$

\begin{figure}[h]
\centering
\includegraphics[width=60mm]{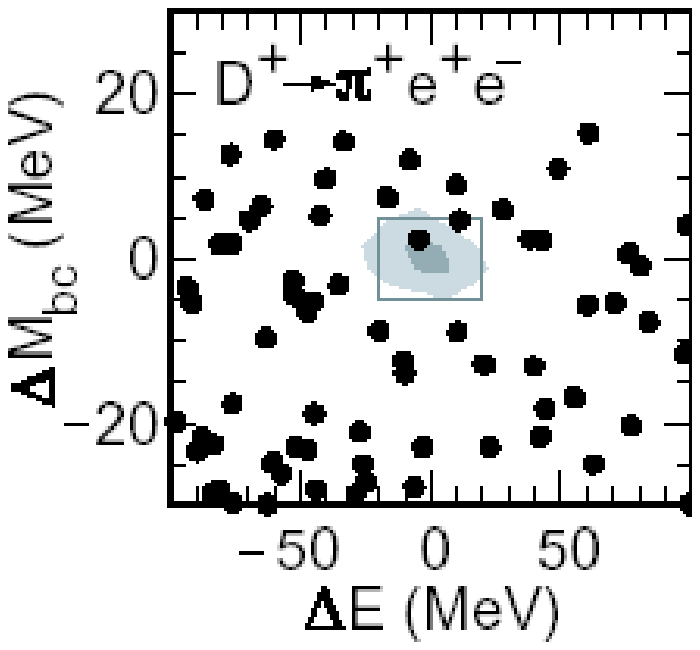}
\includegraphics[width=60mm]{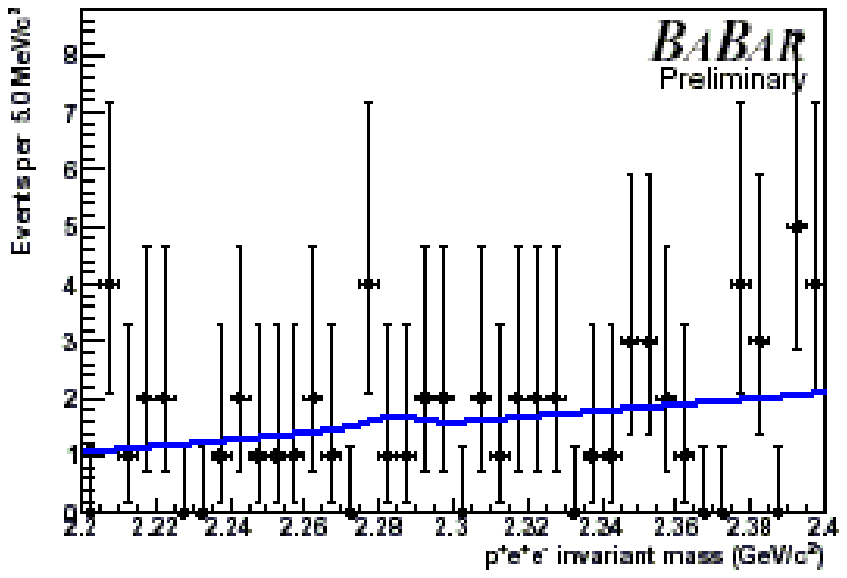}
\includegraphics[width=60mm]{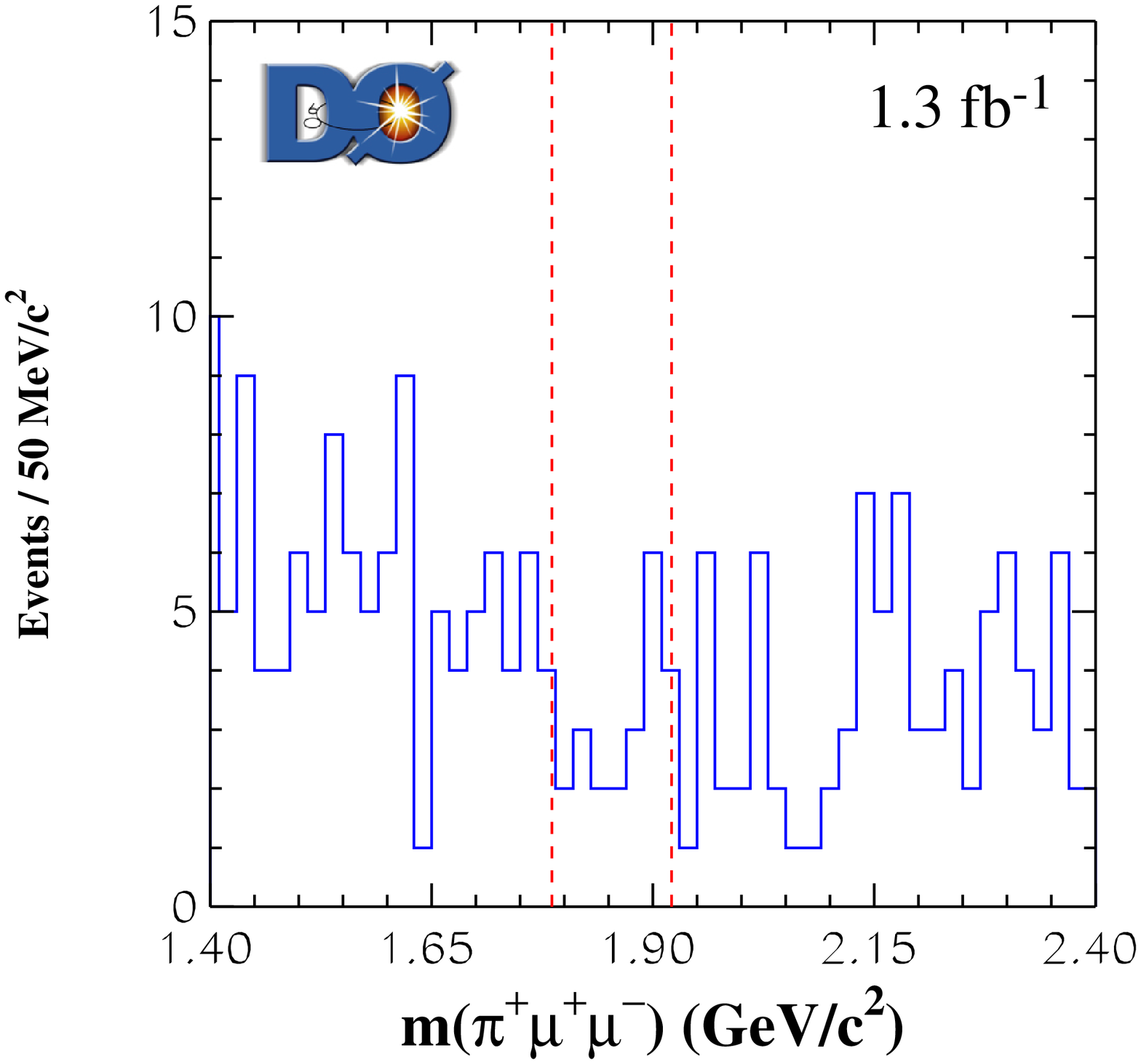}
\caption{Results of the search for $c\rightarrow u l^+l^-$.  The top figure is the beam constrained mass versus beam energy difference from CLEO in the $D^+\rightarrow \pi^+ e^+e^-$  channel.  The middle figure is the $\Lambda_c\rightarrow p e^+e^-$ invariant mass from BaBar.  The lower figure is the $D^+\rightarrow \pi\mu^+\mu^-$ invariant mass distribution from D\O .} \label{fig:ull}
\end{figure}

The missing energy resolution of the D\O  detector does not allow them to 
veto semileptonic events where the neutrinos typically carry away a few GeV
 of energy and the long lived backgrounds from semileptonic charm and b hadron
 decay are essentially irreducible.  However the much more dominant background
 is from light quark and Drell-Yann production that can be removed using flight
 length significance, vertex quality, and topological requirements and attempts 
are made to optimize the analysis for both direct $D$ meson production and $D$ 
mesons produced in $B$ meson decay.  Background reduction based on these variables
 allow D\O  to set the best $90\%$ upper limit in the dimuon channel of 
$${\cal B}(D^+\rightarrow \pi^+ e^+e^-) < 3.9\times 10^{-6} \hskip 1mm\rm (D\slash{O}).$$
The results are shown in Fig.~\ref{fig:ull}.
Since many scenarios of new phenomena predict different rates of excess in the 
dimuon and dielectron channels, its encouraging that together BaBar and D\O 
can cover both channels.

In conclusion, the last round of results in rare charm decays is producing 
precision measurements of $D_s$ annihilation branching fractions.  The combination
 of statistical power and results in both the $\tau\nu$ and $\mu\nu$ channel 
may help add to knowledge recently gained from measurements of
 $B^+\rightarrow\tau\nu$, $B\rightarrow D^*\tau\nu$ and $t\rightarrow b\tau\nu$.

The last round of results has also pushed limits on neutral annihilation and 
radiative decay from the $10^{-5}$ level to the $10^{-6}$ level with much of 
the data currently on tape yet to be analyzed.  A complete analysis of the full 
B factory and Tevatron data sets as well as data at a super B factory and LHCb 
should push these results to the 10$^{-7}$ level and hopefully yield an anomalous excess.

\begin{acknowledgments}
I would like to thank the organizers for an excellent conference on a beautiful campus.  
I would also like to thank Paoti Chang, Jim Olsen, and Brian Peterson for help with the
 $B$ factory results. I would also like to acknowledge the papers of Burdman, Golowich,
 Hewett, and Pakvasa as well as those of Fajfer, Prelovsek, and Singer that played an 
important role in motivating these experimental studies. 
It only takes one person per experiment to continue a healthy rare charm decay program.
\end{acknowledgments}

\bigskip 

\end{document}